\documentstyle[aps,multicol,eqsecnum,epsf]{revtex}
\begin{document}
\draft
\title{Avalanches and the Directed Percolation Depinning Model:
Experiments, Simulations and Theory}

\author{L.~A.~N. Amaral,$^1$ A.-L. Barab\'asi,$^{1, \dagger}$ S.~V.
Buldyrev,$^1$ S.~T. Harrington,$^1$ S. Havlin,$^{1,2}$ \\ R.
Sadr-Lahijany,$^1$ and H.~E.  Stanley$^1$}

\address{$^1$Center for Polymer Studies and Dept. of Physics,
Boston University, Boston, MA 02215 USA \\ $^2$Minerva Center and
Dept. of Physics, Bar-Ilan University, Ramat Gan, Israel}

\date{\today}

\maketitle

\begin{abstract}

  We study the recently-introduced directed percolation depinning
(DPD) model for interface roughening with quenched disorder for which
the interface becomes pinned by a directed percolation (DP) cluster
for $d = 1$, or a directed surface (DS) for $d > 1$.  The mapping to
DP enables us to predict some of the critical exponents of the growth
process.  For the case of $(1+1)$ dimensions, the theory predicts that
the roughness exponent $\alpha$ is given by $\alpha = \nu_{\perp} /
\nu_{\parallel}$, where $\nu_{\perp}$ and $\nu_{\parallel}$ are the
exponents governing the divergence of perpendicular and parallel
correlation lengths of the DP incipient infinite cluster.  The theory
also predicts that the dynamical exponent $z$ equals the exponent
$d_{\rm min}$ characterizing the scaling of the shortest path on a
isotropic percolation cluster.  For the case of $(1+1)$ dimensions,
our simulations give $\nu_{\parallel} = 1.73 \pm 0.02$, $\alpha =
0.63\pm0.01$, and $z=1.01\pm0.02$, in good agreement with the theory.
For the case of $(2+1)$ dimensions, we find $\nu_{\parallel} = 1.18
\pm 0.1$, $\alpha = 0.48\pm0.03$, and $z=1.13\pm0.03$, also in accord
with the theory.  For higher dimensions, $\alpha$ decreases
monotonically but does not seem to approach zero for any dimension
calculated ($d \le 6$), suggesting that the DPD model has no upper
critical dimension for the static exponents.  On the other hand, $z$
appears to approach $2$ as $d \rightarrow 6$, as expected by the
result $z = d_{\rm min}$, suggesting that $d_c = 6$ for the dynamics.
We also perform a set of imbibition experiments, in both $(1+1)$ and
$(2+1)$ dimensions, that can be used to test the DPD model.  We find
good agreement between experimental, theoretical and numerical
approaches.  Further, we study the properties of avalanches in the
context of the DPD model.  In $(1+1)$ dimensions, our simulations for
the critical exponent characterizing the duration of the avalanches
give $\tau_{\rm surv} = 1.46\pm0.02$, and for the exponent
characterizing the number of growth cells in the interface $\delta =
0.60\pm0.03$.  In $(2+1)$ dimensions, we find $\tau_{\rm surv} =
2.18\pm 0.03$, and $\delta = 1.14\pm 0.06$.  We relate the scaling
properties of the avalanches in the DPD model to the scaling
properties for the self-organized depinning (SOD) model, a variant of
the DPD model.  We calculate the exponent characterizing the
avalanches distribution $\tau_{\rm aval}$ for $d = 1$ to $d = 6$, and
compare our results with recent theoretical predictions.  Finally, we
discuss a variant of the DPD model, the ``gradient DPD model'', in
which the concentration of pinning cells increases with height.  We
perform a set of experiments in $(1+1)$ dimensions that are well
described by the gradient DPD model.

\end{abstract}

\pacs{PACS numbers: 47.55.Mh 68.35.Fx}

\begin{multicols}{2}

\section{Introduction}

  Recently the growth of rough interfaces has witnessed an explosion
of theoretical, numerical, and experimental studies, fueled by the
broad interdisciplinary aspects of the subject
\cite{vfk,krug,meakin,hhz,alb,grs}.  Applications can be so diverse as
imbibition in porous media, fluid--fluid displacement, bacterial
colony growth, fire front motion, and the motion of flux lines in
superconductors
\cite{stokes,rubio,horv1,horv2,wong,zzhang,Havlin,Buldyrev,Buldyrev1,Buldyrev2,Amaral}.

  In general, a $d$-dimensional self-affine interface, described by a
single-valued function $h(x,t)$, evolves in a $(d+1)$-dimensional
medium.  Usually, some form of disorder $\eta$ affects the motion of
the interface leading to its roughening.  Two main classes of disorder
have been discussed in the literature.  The first, called thermal or
``annealed'', depends only on time.  The second, referred to as
``quenched'', is frozen in the medium.  Early studies focused on
time-dependent uncorrelated disorder as being responsible for the
roughening.  Here, we focus on the effect of quenched disorder on the
growth.

  The roughening process can be quantified by studying the {\it
global\/} interface width $W$
\begin{equation}
W(L,t) \equiv \left \langle \left ( \overline{h^2({\bf x},t)} -
\overline{h({\bf x},t)}^2 \right ) ^{1/2} \right \rangle,
\label{wid}
\end{equation}
where $L$ is the system size, the bar denotes a spatial average, and
the brackets denote an average over realizations of the disorder.  The
study of discrete models \cite{vis,family,mrsc,kk} and continuum
growth equations \cite{EW,KPZ} leads to the observation that during
the initial period of the growth, i.e. for $t \ll t_{\times}(L)$, the
width grows with time as
\begin{equation}
W(t) \sim t^{\beta} \qquad [t \ll t_{\times}],
\label{bet}
\end{equation}
where $\beta$ is the {\it growth\/} exponent.  For times much larger
than $t_{\times}$ the width saturates to a constant value.  It was
observed that the saturation width of the interface $W_{\rm sat}$
scales with $L$ as
\begin{equation}
W_{\rm sat} \sim L^{\alpha} \qquad [t \gg t_{\times}],
\label{wsat}
\end{equation}
where $\alpha$ is the {\it roughness\/} exponent.  The dependence of
$t_{\times}$ on $L$ allows the combination of (\ref{bet}) and
(\ref{wsat}) into a single scaling law \cite{vis}
\begin{mathletters}
\begin{equation}
W(L,t) \sim L^{\alpha}~ f_1( t / t_{\times} ),
\end{equation}
where
\begin{equation}
t_{\times} \sim L^{z}.
\end{equation}
\label{scl1}
\end{mathletters}
\noindent
Here $z = \alpha / \beta$ is the {\it dynamical\/} exponent, and
$f_1(u)$ is a universal scaling function that grows as $u^{\beta}$
when $u \ll 1$, and approaches a constant when $u \gg 1$.

  An alternative way of determining the scaling exponents is to study
the {\it local\/} width $w$ in a window of observation of length $\ell
< L$.  The scaling law (\ref{scl1}), and the fact that the interface
is self-affine, allow us to conclude
\begin{mathletters}
\begin{equation}
w(\ell,t) \sim \ell^{\alpha}~ f_2( \ell / \ell_{\times} ),
\end{equation}
where
\begin{equation}
\ell_{\times} \sim t^{1/z} \qquad [t \ll t_{\times}],
\end{equation}
or
\begin{equation}
\ell_{\times} \sim L \qquad [t \gg t_{\times}].
\end{equation}
\label{scl2}
\end{mathletters}
\noindent
Here $f_2(u)$ is a universal scaling function that decreases as
$u^{-\alpha}$ when $u \gg 1$ and approaches a constant when $u \ll 1$.

  The simulation of discrete models \cite{vis,family,mrsc,kk} gives
exponents in agreement with the predictions of phenomenological
continuum approaches, such as the Edwards-Wilkinson (EW) equation
\cite{EW} and the Kardar-Parisi-Zhang (KPZ) equation \cite{KPZ}.  However,
experimental studies find exponents significantly larger than the
predictions of theory --- for example, for $(1+1)$ dimensions, Refs.
\cite{EW,KPZ} predict $\alpha=1/2$ but experiments show $\alpha \simeq
0.6-1.0$
\cite{stokes,rubio,horv1,horv2,wong,zzhang,Havlin,Buldyrev,Buldyrev1,Buldyrev2,Amaral}.
Although various explanations were proposed --- long-range
correlations \cite{Medina}, power-law distribution \cite{Zhang} for
the disorder, or coupling of the interface to impurities
\cite{Barabasi} --- it is currently accepted that {\it quenched\/}
disorder plays an essential role in those experiments
\cite{Havlin,Buldyrev,Buldyrev1,Buldyrev2,Amaral,Tang,robbins,rob2,kessler,parisi,leschhorn,dong,jullien,alstrom,csahok,Amaral+Makse,Kardar,makse,theory1,theory2}.

  The presence of quenched disorder allows an interesting analogy with
critical phenomena.  The continual motion of the interface requires the
application of a driving force $F$.  There exists a critical value $F_c$
such that for $f < F_c$, the interface will become pinned by the
disorder after some finite time.  For $F > F_c$ the interface moves
indefinitely with a constant velocity $v$.  This means that the motion
of driven rough interfaces in disordered media can be studied as a phase
transition --- called the depinning transition.  The velocity of the
interface $v$ plays the role of the {\it order parameter}, since as $F
\rightarrow {F_c}^+$, $v$ vanishes as
\begin{equation}
v \sim f^{\theta},
\label{vel}
\end{equation}
where $\theta$ is the {\it velocity\/} exponent, and $f \equiv
(F-F_c) / F_c$ is the {\it reduced force} (Fig \ref{phases}).

  For $F \rightarrow {F_c}^+$, large but finite regions of the interface
are pinned by the disorder.  At the transition, the characteristic
length $\xi$ of these pinned regions diverges,
\begin{equation}
\xi \sim f^{-\nu},
\label{corl}
\end{equation}
where $\nu$ is the {\it correlation length\/} exponent.

  Several models in which quenched disorder plays an essential role
have been proposed recently
\cite{Havlin,Buldyrev,Buldyrev1,Buldyrev2,Amaral,Tang,robbins,rob2,kessler,parisi,leschhorn,dong,jullien,alstrom,csahok,Amaral+Makse,Kardar,makse,theory1,theory2}.
For one class of models \cite{Havlin,Buldyrev,Tang}, in $(1+1)$
dimensions, $\alpha$ can be obtained exactly by mapping the interface,
at the depinning transition onto {\it directed percolation\/} (DP).
In higher dimensions the interface can be mapped to {\it directed
surfaces\/} (DS) \cite{Buldyrev1}.  In $(1+1)$ dimensions, DP and DS
are equivalent.  We refer to this class of models as the {\it directed
percolation depinning\/} (DPD) universality class.

  Recent numerical studies \cite{Amaral+Makse}, confirmed by
analytical arguments \cite{Kardar}, showed that this class of models
can be described by a stochastic differential equation of the KPZ type
\cite{csahok}
\begin{equation}
\frac{\partial h}{\partial t} = F + \nabla^2 h + \lambda
(\nabla h)^2 + \eta({\bf x},h),
\label{qkpz}
\end{equation}
where $\eta({\bf x},h)$ represents the quenched disorder, and the
coefficient $\lambda$ of the nonlinear term diverges at the depinning
transition \cite{Amaral+Makse}.  This equation was originally proposed
in the context of interface roughening in the presence of quenched
disorder in Refs. \cite{csahok}.  The numerical integration of
(\ref{qkpz}) yielded exponents in agreement with the calculations for
the models in the DPD universality class \cite{csahok}.

  For a number of different models
\cite{robbins,rob2,kessler,parisi,leschhorn,dong,makse} belonging to a
second universality class --- referred to as {\it isotropic growth\/}
--- we have either $\lambda = 0$ or $\lambda \rightarrow 0$ at the
depinning transition \cite{Amaral+Makse}.  So, near the depinning
transition, they can be described by an equation of the EW type with
quenched disorder \cite{bruins}
\begin{equation}
\frac{\partial h}{\partial t} = F + \nabla^2 h + \eta({\bf x},h).
\label{qew}
\end{equation}
This equation has been studied by means of the functional
renormalization group \cite{theory1,theory2}, yielding $\alpha =
\epsilon / 3$, $\nu = 1 / (2 - \alpha)$, and $z = 2 - 2 \epsilon / 9$,
where $\epsilon = 4 - d$.

  When $F \gg F_c$, the size of the pinned regions in the interface
$\xi$ decreases to values much smaller than the system size $L$.  For
length scales $\ell$ larger than $\xi$, the quenched disorder becomes
irrelevant and time-dependent noise dominates the roughening process.
This means that for $\ell \gg \xi$ we should recover the results of
either the EW or the KPZ equation with annealed noise (depending on
the absence or presence of nonlinear terms).  This behavior has been
observed both in experiments \cite{stokes,rubio,horv1,horv2,wong,zzhang} and
in simulations of discrete models
\cite{Havlin,Buldyrev,Buldyrev1,Tang,robbins,rob2,kessler,parisi,leschhorn,dong,makse}.

  The DPD model, discussed in this paper, was introduced in
Refs. \cite{Havlin,Buldyrev} to explain a set of simple imbibition
experiments --- a somewhat different model, belonging to the same
universality class, was independently introduced in Ref. \cite{Tang}.
In these experiments a colored suspension (coffee or ink) imbibes a
sheet of paper, in the $(1+1)$-dimensional case --- or a porous,
spongy-like brick, in the $(2+1)$-dimensional case
\cite{Havlin,Buldyrev,Buldyrev1,Buldyrev2,Amaral}.  The
experimentally-measured roughness exponents are in good agreement with
the predictions of the DPD models
\cite{Havlin,Buldyrev,Buldyrev1,Tang}.  However, a number of
experimental features cannot be explained by this model.  For example,
in the experiments, the saturation width and the average height of the
pinned interface depend on the rate of evaporation, which is not taken
into account in the DPD model.  We will also discuss a variant of the
DPD model that enables us to explain the experimental results in terms
of the effect of evaporation \cite{Amaral}.

  A self-organized variant of the DPD model has also been studied
\cite{Havlin,Sneppen} .  In this model the growth proceeds by
avalanches, whose properties are not only of interest for the study of
interface roughening \cite{sj,lt,olami,maslov}, but also for other
fields, including biological evolution in ecological systems
\cite{soc,biol1,biol2,biol3}.

  The paper is organized as follows.  In Sec. II we describe a set of
imbibition experiments, in both $(1+1)$ dimensions and $(2+1)$
dimensions, that allow us to study the scaling properties of interface
roughening in media with quenched disorder.  In Sec. III we discuss
the DPD model, and calculate its relevant exponents.  We analyze its
mapping to DP for one-dimensional interfaces (and DS, for higher
dimensions), and its connection to the universality class of the KPZ
equation with quenched disorder.  In Sec. IV, we define avalanches and
discuss their scaling properties and their relation to the
self-organized depinning (SOD) model.  In Sec. V, we describe a new
set of experiments that probe the ``reason'' for the pinning of the
interface.  We link this pinning to a gradient in the driving force
generated by evaporation and the fluid properties.  We find that the
scaling of the interface width changes with the evaporation rate and
is characterized by an exponent $\gamma$.  We then introduce a variant
of the DPD model that explains the experimental features and present
the results of calculations for this model.  Finally, in Sec. VI we
summarize the main results of the paper.  Some of the results were
presented in a preliminary form in several conference proceedings
\cite{Havlin,Buldyrev1,Buldyrev2}.

\section{Imbibition Experiments}

  In this section we describe a set of experiments --- originally
proposed in Refs. \cite{Havlin,Buldyrev,Buldyrev1} --- that allow us
to study the scaling properties of a rough interface moving in a
disordered medium.  In these experiments, a colored suspension imbibes
an absorbing material, in $(1+1)$ dimensions a sheet of paper, and in
$(2+1)$ dimensions a spongy brick or a paper roll.

\subsection{The Case of $(1+1)$ Dimensions}

  In the $(1+1)$-dimensional experiments, a sheet of paper of edge 20
cm is dipped into a reservoir filled with a colored suspension: coffee
(see Fig. \ref{exper}).  A wet region starts to grow and a rough
interface between dry and wet regions, the wetting front, propagates
in the paper.

  Although the experiments are quite simple to describe, the prediction
of the scaling properties of the wetting front from first principles
is very complex. At microscopic length scales, paper is an extremely
disordered substance, formed by long fibers that are randomly
distributed and connected.  The wetting fluid propagates along the
fibers of the paper due to capillary forces.

  The advance of the wetting front depends on many factors: fluid
density (which depends on the evaporation rate), suspension viscosity
and density, gravity, temperature, size of the holes between fibers,
etc.  Some of these factors (evaporation rate, gravity, fluid density)
modify the {\it effective\/} value of the driving force that leads to
the advance of the wetting front.  Other factors, like the size of the
holes between fibers, determine the pinning force that opposes the
growth of the wet region.

  Certain regions of the paper can locally pin the advance of the wet
region.  If such a region spans the entire system the interface can
become globally pinned, so the advance of the wetting front is stopped
\cite{surf}.  As the interface departs from the water source,
evaporation is constantly decreasing the fluid density, making it more
and more difficult for the wetting front to advance.  Eventually, a
critical height $h_c$ is reached for which the fluid pressure becomes
such that the wetting of the regions above becomes impossible.

  When the interface becomes pinned, we digitize the rough boundary
between wet (colored) and dry (uncolored) areas.  In
Fig. {\ref{wexp2d}, we plot the local width $w$ (averaged over $10$
experiments) as a function of the window of observation $\ell$ for the
digitized experimental interface.  We observe a power-law scaling for
nearly $1.5$ decades.  The fit to a power law, supported by
consecutive slope analysis, results for the ``pinned phase'' in
\begin{equation}
\alpha_{\rm exp}^{pp} = 0.63 \pm 0.04 \qquad [d = 1].
\label{ex01}
\end{equation}

  We repeated the experiments but this time not allowing enough time
for the interface to get pinned.  We then digitized the interface and
plotted $w$ against $\ell$ (Fig. \ref{wexp2d}).  The best fit to a
power law results for the ``moving phase'' in
\begin{equation}
\alpha_{\rm exp}^{mp} = 0.73 \pm 0.05 \qquad [d = 1].
\label{ex02}
\end{equation}
Although the error bars show that the two values could be identical, the
analysis of the DPD model in the next section suggests that they should
be different.

\subsection{The Case of $(2+1)$ Dimensions}

  We perform two sets of experiments in $(2+1)$ dimensions.  In the
first, we used a ``Oasis'' brick as the disordered medium.  This brick
is made of a spongy-like material and is used by florists to absorb
excess water.  In the second set of experiments, we used a paper-towel
roll.  As the invading fluid, we tested several suspensions and found
the most appropriate to be ``Bingo'' ink because of its high viscosity
and good coloring.

  To insure good absorption of fluid by the spongy brick or the paper
roll, we placed them over small ball bearings.  In both sets of
experiments, we added ink gradually and periodically, in order to
maintain a fairly constant level of ink in the container.

  Unfortunately, we could not be sure if the interface had in fact
become pinned everywhere or not, but we always allowed enough time for
the interface to propagate for several centimeters into the absorbing
media.  After the propagation period, we sliced the ``Oasis'' brick
into longitudinal sections and digitized the rough interface.  For the
paper roll, we selected and digitized several sheets from different
radii (Fig. \ref{exp2pic}).

  We calculated the local width $w$ for 13 slices of the brick and 9
sheets of paper.  In Fig. \ref{exp2wid}, we show the average values of
the local width for the brick and the paper roll.  We find power law
scaling over roughly one decade.  The best fit results in an exponent
\begin{equation}
\alpha_{\rm exp} = 0.52 \pm 0.04 \qquad [d = 2].
\label{ex03}
\end{equation}

  When we compare this result with the calculations for the DPD model
in $(2+1)$ dimensions, we will see that within the error bars is
consistent with both the results for the pinned phase or the moving
phase.  In the $(1+1)$-dimensional experiments we are able to be
certain that the interface is completely pinned when we digitize it.
On the other hand, in the case of $(2+1)$ dimensional the interface
may still be moving when we interrupt the experiments.  In fact, in
the case of $(2+1)$ dimensions, it is not clear if evaporation can
have any role in the pinning of the interface.

\section{The DPD Model}

  The imbibition experiments described in the previous section are too
complex to be described from first principle theories.  For this
reason it is convenient to develop a model that captures the most
important features of the experiments.  With this approach we lose the
possibility of exactly predicting the form of the rough interface
between the wet and dry regions.  Instead, we will be able to explain
the scaling properties of the interface.  The DPD model was initially
introduced in Refs. \cite{Havlin,Buldyrev} (and independently in
Ref. \cite{Tang}) with such intentions.  The model was studied in more
detail in Refs.
\cite{Buldyrev,Buldyrev1,Buldyrev2,Amaral,Tang}.

  In this section, we will describe the DPD model in $(1+1)$
dimensions; the generalization for higher dimensions is immediate.  We
then proceed to discuss its most important properties.  In the next
section we discuss the concept of avalanches in the DPD model and
study their scaling properties.

\subsection{Description of the Model}

  Let us consider a square lattice of edge $L$ with periodic boundary
conditions along the direction of that edge.  To each cell $i$ of the
lattice we assign an uncorrelated random number, the disorder
$\eta_i$, with magnitude uniformly distributed in the interval
$[0,1]$.  The role of $\eta_i$ is to model the random pinning forces
generated by the disorder.  We compare the random pinning forces
$\eta_i$ in the lattice with the driving force $F$, where $0 \le F \le
1$.  If the pinning force at a certain cell $\eta_i$ is larger than
the driving force, the cell is labeled ``blocked'', otherwise it is
labeled ``unblocked''.  Thus a cell is blocked with a probability
\begin{equation}
p = 1 - F.
\label{prob}
\end{equation}

  Since the model was developed to study imbibition, we will refer to
the growing, invading, region as ``wet'', and to the invaded region as
``dry''.  At time $t=0$, we wet all cells in the bottom row of the
lattice.  Then, we select a column at random \cite{expl0}, and wet all
dry {\it unblocked cells\/} in that column that are nearest-neighbors
to a wet cell.  To obtain a single-valued interface, we impose the
auxiliary rule that all dry {\it blocked cells\/} below a wet cell
become wet as well \cite{explain0} (see Fig. \ref{model}).  We refer
to this rule as {\it erosion of overhangs}.  The time unit is defined
as $L$ growth attempts.

  We define pinning clusters to be any group of blocked cells that are
connected through nearest or next-nearest-neighbor blocked cells.  Any
pinning cluster whose linear size $\xi$ is smaller than the system size
cannot prevent the advance of the interface.  In fact, any pinning
cluster that does not span the system will eventually become surrounded
the invading fluid, since the invading front can move around finite
``obstacles'', and the erosion of overhangs rule implies that after
being surrounded the pinning cluster becomes wet.

\subsection{Connection to DP}

  Figure \ref{pinmodel} demonstrates that the advance of the wet region
can only be pinned by a directed path of blocked cells that spans the
lattice.  By {\it directed\/} path we mean a connected path of blocked
cells that does not turn back.  Note that if a path turns back, the the
part of the path turning back eventually would become surrounded and
hence wet.

  Such pinning clusters are branches of a DP cluster, a fact which
enables us to map the scaling properties of the pinned interface to the
scaling properties of DP clusters \cite{Havlin,Buldyrev,Tang}.  For a
probability of blocked cells smaller than a critical value $p_c$, the DP
clusters are finite.  An infinite cluster is present for $p \ge p_c$.
For the DPD model we find $p_c \simeq 0.470$, consistent with
calculations for DP \cite{perc1,perc2}.  Near $p_c$, the size of DP clusters
is characterized by a longitudinal (parallel) correlation length
$\xi_{\parallel}$ and a transverse (perpendicular) correlation length
$\xi_{\perp}$ that when $p \rightarrow p_c$ diverge as
\begin{equation}
\xi_{\parallel} \sim |p_c-p|^{-\nu_{\parallel}}, \qquad \xi_{\perp} \sim
|p_c-p|^{-\nu_{\perp}}.
\label{xi}
\end{equation}
The parallel and perpendicular correlation length exponents for DP
clusters have been calculated \cite{EE}, with the results
\begin{equation}
\nu_{\parallel} = 1.733 \pm 0.001, \quad \nu_{\perp} = 1.097 \pm 0.001
\qquad [d = 1].
\label{exp0}
\end{equation}

\subsection{Scaling Properties}

  The mapping of the {\it pinned interface\/} to DP enables us to
estimate the static exponents of this problem from the characteristic
exponents of DP clusters.  The characteristic length $\xi$ of the
pinned regions must be of the order of $\xi_{\parallel}$, so we can
identify the exponent $\nu$ to be
\begin{equation}
\nu = \nu_{\parallel}.
\label{nu}
\end{equation}

  The global width $W_{\rm sat}$ of the pinned interface should scale
as $\xi_{\perp}$, since its advance is blocked by a DP path.  On the
other hand, $\xi_{\parallel}$ must be larger than the system size $L$
for the interface to become pinned, from which follows
\cite{Havlin,Buldyrev,Tang}
\begin{equation}
W_{\rm sat} \sim \xi_{\perp} \sim \xi_{\parallel}^{\nu_{\perp} /
\nu_{\parallel}} \sim L^{\nu_{\perp} / \nu_{\parallel}} \qquad
[\xi_{\parallel} \ge L].
\label{eq100}
\end{equation}
Comparing with (\ref{wsat}), we conclude that the roughness exponent
is given in terms of the correlation exponents for DP,
\begin{equation}
\alpha = \nu_{\perp}/\nu_{\parallel}.
\label{alp}
\end{equation}
Substituting (\ref{exp0}) into (\ref{alp}), we predict
\begin{equation}
\alpha = 0.633 \pm 0.001 \qquad [d = 1],
\label{alpxx}
\end{equation}
in good agreement with the experiments and the simulations.  For the
case $\xi_{\parallel} \ll L$, we obtain the KPZ result with annealed
disorder for length scales $\ell$ such that $\xi_{\parallel} \ll \ell
\ll L$.

  If $p \ge p_c$, the interface becomes pinned after some finite time.
However, for $p < p_c$, the DPD model gives rise to an interface which
propagates with a constant, non-zero, velocity.  In Sec. I, we
discussed how near the depinning transition the velocity of the
interface scales as a power law.  To determine the velocity exponent
let us consider the following argument.  Near the depinning transition
most of the interface is pinned, except for a few regions.  The growth
occurs by the lateral propagation of the growing regions through the
system.  The characteristic time required for this propagation is of
the order of $t_{\times}$.  During this process the interface advances
from one blocking path to the next, the distance advanced is typically
of the order of $\xi_{\perp} \sim W_{\rm sat}$.  Using (\ref{wsat}),
(\ref{scl1}), (\ref{corl}) and (\ref{xi}) we obtain
\begin{equation}
v \sim \xi_{\perp} / t_{\times} \sim \xi_{\parallel}^{\alpha} /
\xi_{\parallel}^{z} \sim \xi_{\parallel}^{\alpha - z} \sim
f^{-\nu_{\parallel} (\alpha - z)}.
\label{eq101}
\end{equation}
Upon comparison with (\ref{vel}), we conclude that
\begin{equation}
\theta = \nu_{\parallel} (z - \alpha).
\label{thet}
\end{equation}
This relation can also be derived in a different way
\cite{theory1,theory2}.

  Reference \cite{Amaral+Makse} showed that for the DPD model, the
coefficient $\lambda$ of the nonlinear term of Eq. (\ref{qkpz})
diverges at the depinning transition
\begin{equation}
\lambda \sim f^{-\phi}.
\label{lamb}
\end{equation}
The new exponent $\phi$ can be linked to the other critical exponents
that characterize the depinning transition \cite{Kardar}
\begin{equation}
\phi = \nu_{\parallel} (2 - \alpha - z).
\label{phi1}
\end{equation}
This prediction is in good agreement with the calculations of
Ref. \cite{Amaral+Makse} for $d \le 2$.

  The mapping to DP is not unique to the DPD model, but is a general
feature of a large class of models
\cite{Buldyrev2,Tang,parisi,Kardar}.  In $(1+1)$ dimensions the
agreement between the estimates of the exponents from the mapping to
DP and the numerical simulations is quite good.

  An interesting problem, still unsolved, is the situation for the
moving interface.  Measurements of the local width $w$ in function of
the window of observation $\ell$ lead to an effective exponent
$\alpha^{mp} \simeq 0.72$ \cite{Buldyrev1,Tang}.  However, it is possible
that the interface in the moving regime might not be self-affine
\cite{Tang}.

\subsection{Higher Dimensions}

  In higher dimensions, the mapping to DP is no longer possible, since
the DP clusters have many holes making it impossible for them to pin the
interface.  In this case the interface can only be pinned by a DS which
is a $d$-dimensional hypersurface embedded in a $(d+1)$-dimensional
space.  The DS are self-affine, and are ``directed'' in the sense that
they do not have overhangs.  In $(1+1)$ dimensions, DS reduces to DP.

  DS is a new problem about which little is known. Near the critical
probability --- i.e., the probability for which their size diverges ---
they can be characterized by two correlation lengths that scale as
(\ref{xi}).  All the relations derived in the previous subsection for
the scaling properties of the interface are still valid for higher
dimensions, but $\nu_{\parallel}$ and $\nu_{\perp}$ are now the
correlation length exponents for DS.

  We thoroughly studied the DPD models for dimensions up to $(6+1)$.
The results for most of the exponents of interface roughening in
disordered media are presented in Table \ref{expdim} (see also
Fig. \ref{dpalp}).

  For the case of $(2+1)$ dimensions, we also calculated the roughness
exponent for the moving interface and obtained
\begin{equation}
\alpha^{mp} = 0.52 \pm 0.03 \qquad [d = 2],
\label{eq90}
\end{equation}
in good agreement with the value (\ref{ex03}) obtained in the
experiments.

\subsection{The Dynamical Exponent $z$}

  In the previous subsections, we derived scaling relations linking the
exponents characterizing the static properties of rough interfaces in
the presence of quenched disorder using the mapping of the pinned
interface to DP ($d = 1$) or DS ($d > 1$).  However, the dynamics of the
roughening was not derived, leaving us with an unknown exponent: $z$.
In this subsection, we determine $z$ from the study of how a
perturbation, caused by a single unblocked cell, propagates over the
interface \cite{habhs}.

  To study the propagation of correlations in the system it is better
to change the initial conditions in the model and to make the invasion
of the cells nearest neighbors to the wet region in parallel
(Fig. \ref{avalproj}).  If we start the simulation at time $t = 0$
with a single unblocked cell in front of the interface we can follow
the propagation of correlations simply by monitoring the longitudinal
dimension of the invaded region.  At each time step, a certain set of
cells becomes invaded. In analogy with invasion percolation, we call
this set of cells the {\it percolation shell}.  We refer to the
longitudinal and perpendicular components of the average radius of
gyration of the percolation shell by $r_{\parallel}$ and $r_{\perp}$,
respectively.  From (\ref{scl2}) we see that
\begin{equation}
t \sim r_{\parallel}^z.
\label{rpar}
\end{equation}

  For the case $d = 1$, all shells are confined between the old directed
path that spans the system at $t=0$ and a new pinning path that will
block the growth after some time. The region between these two paths is
effectively one-dimensional, since the vertical distance between them
scales as $\xi_{\perp}$, and $\nu_{\perp} < \nu_{\parallel}$ implies
$\xi_{\perp} / \xi_{\parallel} \rightarrow 0$ as $f \rightarrow 0$.

  For any cell on the interface that becomes wet at time $t$, we can
find the cell from which it was invaded at the previous time step, and
recreate the sequence of invasion events that leads from the initial
cell to any given cell on the interface (Fig. \ref{tx1}). The
trajectory of this sequence follows the upper pinning path and is
effectively one dimensional. Its length $\ell$ scales as its average
end-to-end distance $r_{\parallel}$. On the other hand, $\ell$ is
equal to the time $t$ needed to reach the end of the path.  Hence $t
\sim r_{\parallel}$, and from (\ref{rpar}) we conclude that $z=1$.
This result is supported by our simulations (Table \ref{tab1}).

  For the case $d > 1$, we must consider the region bounded by two
self-affine DS. This region is effectively $d$-dimensional, since
$\xi_{\perp} / \xi_{\parallel} \rightarrow 0$. Hence, the shortest
path leading from the initial point to any point of this region is
effectively confined to a $d$-dimensional hyperplane
(Fig. \ref{tx2}). This shortest path has to avoid blocked cells in
this hyperplane, {\it as does\/} the shortest path of isotropic
percolation.  For isotropic percolation it is known that the length of
the shortest path $\ell$ scales with the Euclidean end-to-end distance
$r$ as $\ell \sim r^{d_{\rm min}}$ \cite{perc1}.  The similarity
between the geometrical properties of the paths in DPD and isotropic
percolation leads us to propose
\begin{equation}
z = d_{\rm min}.
\label{zdmin}
\end{equation}
To test the argument leading to (\ref{zdmin}), we perform simulations
for both DPD and percolation for $d = 1$ to $d = 6$.  We present our
results for both $z$ and $d_{\rm min}$ in Table \ref{tab1}
\cite{fitz}.

  It is well known that for isotropic percolation the upper critical
dimension is $d_c = 6$, i.e. for $d > d_c$ the mean field result,
$d_{\rm min} = 2$, becomes exact \cite{perc1}.  This suggests an upper
critical dimension, $d_c + 1 = 7$, for the {\it dynamics\/} of the DPD
models which are in the universality class of Eq. (\ref{qkpz}), and
that $z = 2$ for $d + 1 \ge 7$.

  Since the dynamics of Eq. (\ref{qkpz}) and the models in the DPD
universality class are connected to isotropic percolation, while the
static properties are mapped to DP or DS, it is possible that the
upper critical dimension determined in this study {\it may be valid
only for the dynamics}.  It is also possible that $d_c$ for the static
properties may not exist, on the basis of the following argument.
When, e.g., a one-dimensional object is embedded in a $d$-dimensional
space, we expect that as we increase $d$ the interactions between the
different parts of the object decrease.  At a certain $d = d_c$, these
interactions can be neglected, and the exponents become those of the
ideal non-interacting case.  In contrast, when the dimension of the
object is not fixed but increases with $d$, as in the case of DS in
which the object is one dimension smaller than the space, we expect to
move away from the non-interacting limit.  In fact, the analytical
solution of the DPD model in the Cayley tree suggest that the upper
critical dimension for the {\it statics\/} might be $\infty$
\cite{Buldyrev2}.

  Thus we see that the DPD model has three {\it independent\/}
exponents: $\nu$, $\alpha$, and $z$.  The static exponents, $\nu$ and
$\alpha$, can be evaluated from the exponents of DP ($d = 1$) or DS ($d
> 1$).  On the other hand, the dynamics of the model are related to
isotropic percolation in $d$ dimensions.  We find that $z$ is equal to
the exponent $d_{\rm min}$ characterizing the scaling of the shortest
path in isotropic percolation.  The model also allows us to calculate
the roughness exponents determined in the experiments both for the
pinned and moving interfaces.  However, at this time, no explanation is
available for why $\alpha$ changes value at the transition nor for the
value of $\alpha^{mp}$.

\section{The Avalanche Mechanism of Interface Motion}

  The study of the scaling properties of self-organized systems is of
great importance for many fields
\cite{sj,lt,olami,maslov,soc,biol1,biol2,biol3}.  In this section we
show how avalanches can be generated in the DPD model, and we study
their properties.  We then relate those avalanches to the SOD variant
and discuss some of the results that have been obtained recently for
the SOD model and compare them with our numerical results.

\subsection{Avalanches in the DPD Model}

  As we discussed above, for $p > p_c$ the growth of the interface for
the DPD model is stopped by the spanning path of a DP cluster in $(1+1)$
dimensions --- or a spanning self-affine DS in $(d+1)$ dimensions.  Even
when the growth is completely stopped, the blocked cells on the
interface may still erode --- but at an {\it infinitesimal\/} rate.
With this assumption, we can remove a blocked cell at random when the
interface is completely stopped, thereby producing an avalanche which
eventually will die out when the front reaches a second pinning
hypersurface (Fig. \ref{aval}).

  An alternative way of producing avalanches is to start the growth from
a single wet cell in a row of blocked cells at time $t=0$ (Fig.
\ref{avalproj}).  For $p > p_c$, the growing clusters will eventually
become pinned by a blocking path.  Below $p_c$ most clusters will grow
indefinitely, although some might become pinned by the blocking
surfaces.

  In analogy with conventional percolation, the survival probability
$P_{\rm surv}(t)$ of the clusters will decay as a power law
\cite{rule}
\begin{equation}
P_{\rm surv}(t) \sim t^{-\tau_{\rm surv}} \qquad [t \ll t_{\times}],
\label{tau2}
\end{equation}
where $t_{\times} \sim \xi_{\parallel}^z$.  In Fig. \ref{avaltau},
we show the scaling of $P_{\rm surv}(t)$ for $(2+1)$ dimensions, and the
value of $\tau_{\rm surv}$ is given in Table \ref{expaval}.

  If $p > p_c$, $P_{\rm surv}(t \gg t_{\times})$ approaches zero
exponentially.  If $p < p_c$, $P_{\rm surv}(t \gg t_{\times})$
approaches a constant value $P_{\rm surv}(\infty)$, the probability of
an infinite cluster.  Thus, studying $P_{\rm surv}(t)$ provides a very
accurate method of estimating $p_c$.  We calculated $P_{\rm surv}(t)$
and determined both the exponent $\tau_{\rm surv}$ and $p_c$ for $d = 1$
-- $6$ (Table \ref{expaval}).  Using these high-accuracy determinations
of the critical probability $p_c$ we were able to measure with higher
precision the other critical exponents (Appendix A and Tables
\ref{expdim} and \ref{expaval}).

  The probability distribution of avalanches of ``volume'' $s$,
$P_{\rm aval}(s)$, is calculated as the ratio of the number of
avalanches of volume $s$ to the total number of avalanches. We find
\cite{Buldyrev1}
\begin{equation}
P_{\rm aval}(s)\sim s^{-\tau_{\rm aval}}~ f_3(s / s_o),
\label{densav}
\end{equation}
where $s$ is the number of cells invaded during an avalanche, $s_o
\sim \xi_{\perp}~\xi_{\parallel}^d$ is the characteristic volume of an
avalanche, and $f_3(u)$ is a scaling function that approaches a constant
for $u \ll 1$ and decays exponentially for $u \gg 1$ (Fig.
\ref{avalt1}).

  The exponents $\tau_{\rm surv}$ and $\tau_{\rm aval}$ can be
related, as follows
\begin{equation}
s \sim r_{\perp}^d~r_{\parallel}^d \sim r_{\parallel}^{d +
\alpha} \sim t^{(d + \alpha) / z}.
\label{eq103}
\end{equation}
On the other hand, $P_{\rm aval}(s) ds = P_{\rm surv}(t) dt$. Hence
from (\ref{tau2}) and (\ref{eq103}) follows
\begin{equation}
t^{-\tau_{\rm aval} (d + \alpha) / z}~ t^{(d + \alpha) / z - 1}~
dt = t^{-\tau_{\rm surv}}~ dt.
\label{eq104}
\end{equation}
  From (\ref{eq104}), we find
\begin{equation}
\tau_{\rm aval} = 1 + \frac{z(\tau_{\rm surv} - 1)}{d + \alpha}.
\label{tau3}
\end{equation}

  If $p = p_c$, the dynamical critical exponents $\beta$ and $z$ can be
obtained from the dependence of the avalanche volume $s$ on time.  We
can define a new exponent $\delta$ that characterizes the time
dependence of the number of {\it unblocked cells\/} in the interface ---
i.e., the size of the percolation shell (Fig. \ref{avalproj})
\begin{equation}
n(t) \sim t^\delta.
\label{shell}
\end{equation}
To relate the new exponent $\delta$ to $z$ and $\alpha$, we start by
noticing that the size of the invaded region scales as
\begin{equation}
s \sim r_{\parallel}^{d + \alpha} \sim
{\int}_0^{r_{\parallel}^z} ~n(t) dt \sim r_{\parallel}^
{z (\delta + 1)}.
\label{eq105}
\end{equation}
  From (\ref{eq105}), we obtain
\begin{equation}
\delta = \frac{d + \alpha}{z} - 1.
\label{delt1}
\end{equation}
In $(1+1)$ dimensions, $z=1$, so $\delta=\alpha=\beta$, which agrees
with the simple geometrical picture that the projection of the growing
region scales as the length of the steepest moving terrace, which
scales as the width of the whole system, $W(t)\sim t^\beta$.  In
$(1+1)$ dimensions, we find $\delta = 0.60\pm 0.03$, in agreement with
the above relations.

  The projection of the shell of unblocked cells in the interface
forms a {\it fractal dust\/} (Fig. \ref{avalproj}).  The fractal
dimension of this dust $d_{\rm dust}$ can be related to $\tau_{\rm
surv}$ \cite{Buldyrev2}.  The fractal dimension of the dust must be
the same as the fractal dimension of the ``bubbles'' surrounding the
regions of space confined by two DS.  The distribution of these
confined regions can be linked to the distribution of survival times
for the avalanches.  Thus, we have
\begin{mathletters}
\begin{equation}
P_{\rm aval}(r_{\parallel})~d r_{\parallel} = P_{\rm surv}(t)~d t,
\end{equation}
and
\begin{equation}
r_{\parallel}^{-\tau_{\parallel}}~d r_{\parallel} = t^{-\tau_{\rm
  surv}}~d t.
\end{equation}
\end{mathletters}
Since $t \sim r_{\parallel}^z$, it follows that
\begin{equation}
\tau_{\parallel} - 1 = z (\tau_{\rm surv} - 1).
\end{equation}
On the other hand, it is well known that the fractal dimension of some
dust separated by regions whose size follows a power-law distribution,
$P_{\rm aval}(r_{\parallel}) \sim r_{\parallel}^{-\tau_{\parallel}}$,
is given by
\begin{equation}
d_{\rm dust} = \tau_{\parallel} - 1,
\end{equation}
leading to
\begin{equation}
d_{\rm dust} = z ( \tau_{\rm surv} - 1 ).
\label{ddust}
\end{equation}
Since $d_{\rm dust} < \delta$, the fractal dust is packed in ``moving
blocks''.  These moving blocks behave like quasi-particles, which are
distributed in a fractal way with dimension $d_{\rm dust}$.  This
description is supported by numerical studies of the correlation
function of the dust.

  We can also relate the velocity of the interface to the number of
unblocked cells in the interface, as defined in (\ref{shell}).  The
velocity at each instant can be obtained as the number of unblocked
cells divided by the size of the parallel projection of the invaded
region
\begin{equation}
v \sim n(t) / \xi_{\parallel}^d \sim t^{\delta} / \xi_{\parallel}^d \sim
\xi_{\parallel}^{z \delta - d} \sim f^{-\nu_{\parallel} (z \delta - d)}.
\label{eq106}
\end{equation}
Comparing with (\ref{vel}), we obtain
\begin{equation}
\theta = \nu_{\parallel} (d - z \delta).
\label{thet1}
\end{equation}
As a consistency check, we note that (\ref{thet1}) can also be
obtained by equating (\ref{delt1}) with (\ref{thet}).

  We measured the components of the radius of gyration of the avalanches
in the longitudinal and transverse directions for both $(1+1)$ and
$(2+1)$ dimensions (Fig. \ref{avalnu}).  For $(1+1)$ dimensions we find,
for a system of size $L = 131~072$,
\begin{equation}
\nu_{\parallel}^{\rm aval} = 1.73\pm 0.02,\qquad
\nu_{\perp}^{\rm aval}=1.10\pm 0.02 \qquad [d = 1],
\label{ex32}
\end{equation}
in good agreement with the correlation length exponents of DP given in
Eq. (\ref{exp0}).

  For the case of $(2+1)$ dimensions, we find the correlation length
exponents to be
\begin{equation}
\nu_{\parallel}^{\rm aval} = 1.16 \pm 0.05, \qquad
\nu_{\perp}^{\rm aval} = 0.57 \pm 0.05 \qquad [d = 2],
\label{ex33}
\end{equation}
for a system of linear size $L = 2~048$.  The estimates (\ref{ex33})
predict, from (\ref{alp}), that $\alpha = 0.49\pm 0.1$, in good
agreement with the value obtained from the analysis of the scaling of
the width (see Table \ref{expdim}).

\subsection{Avalanches in the SOD Model}

  A variant of the model discussed so far, in which the growth is made
in an invasion percolation fashion, was introduced in Ref.
\cite{Havlin}.  Subsequently, a similar model was introduced in
Ref. \cite{Sneppen}, and studied in detail in
Refs. \cite{sj,lt,olami,maslov}.  The self-organized depinning (SOD)
model can still be mapped to DP.  In fact, the SOD model is always at
the depinning transition.

  For the SOD model, the cells are not labeled blocked or unblocked.
Instead, we keep in memory the pinning forces at each cell.  The
growth proceeds by the invasion (wetting) of the cell which is a
nearest neighbor to the wet region, and which has the weakest noise
--- i.e. with the smallest random number $\eta$.  Then we erode any
overhangs that might have been formed, i.e.  we wet any dry cell
bellow a wet cell.  The unit time is defined as the volume of the
invaded region, i.e. it takes one time unit to invade one cell.

  At any given instant $t$, the interface is characterized by a set of
noises $\{\eta\}$.  The smallest of these, denoted $\eta_{\rm min}(t)$,
is useful in characterizing the state of the system at that instant.
Reference \cite{lt} argued that every possible configuration of the
interface corresponds to a path on a cluster of sites with values of the
noise greater or equal to $\eta_{\rm min}$.  This fact, and the mapping
to DP implies that $\eta_{\rm min}(t)$ cannot exceed $q_c \equiv 1 -
p_c$, since that would imply the existence of a directed path of
blocked cells spanning the system for a probability smaller than $p_c$.

  The general tendency is for $\eta_{\rm min}(t)$ to increase, until it
reaches $q_c$, but its growth is not monotonic (Fig. \ref{min}).  When
we start the growth from a flat interface the noise has a uniform
distribution between 0 and 1.  This implies that $\eta_{\rm min}(1)$
will be very close to zero.  The interface will then advance to a new
position, and the noise $\eta_{\rm min}(1)$ will be replaced by a new
noise.  Since $\eta_{\rm min}(1)$ is very small then it is very likely
than the new noise will be larger than $\eta_{\rm min}(1)$.  So we see
that $\eta_{\rm min}(2) > \eta_{\rm min}(1)$ with high probability.

  Initially, $\eta_{\rm min}(t)$ will grow monotonically.  However, at
some time $t_o$ all cells in the interface with small noise will have
been eroded, so the probability that $\eta_{\rm min}(t_o+1) <
\eta_{\rm min}(t_o)$ will no longer be irrelevant.  Hence $\eta_{\rm
min}(t)$ can decrease.  During some time interval $\Delta t$, $\eta_{\rm
min}(t)$ will be smaller than $\eta_{\rm min}(t_o)$, which can only
occur in a connected region since all other cells in the interface have
larger noise values.  Therefore we can identify the avalanches with the
regions invaded during the time interval from $t_o$ to $(t_o + \Delta
t)$ \cite{maslov}.

  A different regime is reached when $\eta_{\rm min}(t)$ reaches the
value $q_c$ --- for an infinite system (for a finite system there will
be fluctuations around this value).  Then the interface is just at a
DP path which spans the system.  This implies that for some time steps
the advance will be accomplished by invading cells with $\eta_{\rm
min} < q_c$, until a new DP path is found, for which $\eta_{\rm min}$
is again equal to $q_c$.

  The early stage in which $\eta_{\rm min}$ was smaller than $q_c$ is
referred to as the {\it transient\/} regime; after it ends, the system
is said to be in the {\it critical\/} state.  From then on, the
interface will advance between spanning paths of the DP cluster.  So we
see that the avalanche picture for the SOD model is similar, {\it after
\/} the transient regime, to the avalanche image of the DPD model in
which we unblock a site at random.  This argument is supported by
numerical results \cite{Buldyrev1,Buldyrev2,sj,lt,olami,maslov}.

  The problem with the previous definition of avalanches is that it
is precise only for the critical state.  To remedy this situation, Ref.
\cite{maslov} introduced a modified definition, in which avalanches for
$q < q_c$ are finished whenever $\eta_{\rm min}$ passes the threshold
$q$.

  A different method to define avalanches for $q < q_c$ was proposed
in Refs. \cite{Buldyrev1,Buldyrev2}, whereby the growth is allowed
only until the first time the system has $\eta_{\rm min} = q$.  In
spite of the different definitions and initial conditions, numerical
calculations reveal that both approaches lead to identical exponents,
at least for $d \le 2$.

  Theoretical predictions for the avalanche size distribution were
made in Refs. \cite{sj,lt,olami,maslov}.  We extend an argument
proposed by Olami-Procaccia-Zeitak (OPZ) for $(1+1)$ dimensions
\cite{olami}, to higher dimensions, obtaining
\begin{equation}
\tau_{\rm aval}^{\rm OPZ} = \frac {d + 1}{d + \alpha}.
\label{tau6}
\end{equation}
On the other hand, Maslov and Paczuski (MP) \cite{maslov} predict
\begin{equation}
\tau_{\rm aval}^{\rm MP} = 1+ \frac{d - 1 / \nu_{\parallel} }{d +
\alpha}.
\label{tau7}
\end{equation}

  Our simulations enable us to test the accuracy of both of these
formulas (Table \ref{expaval}).  The OPZ relation (\ref{tau6}) cannot be
correct for all $d$: For $d = 1$, $\tau_{\rm aval}^{\rm OPZ}$ is
slightly below numerical results, but for $d > 1$, (\ref{tau6})
decreases while $\tau_{\rm aval}$ increases.  On the other hand, the MP
relation (\ref{tau7}) predicts values consistent with our simulations.
Moreover, recent work on the Cayley tree \cite{cayley} suggests that the
upper critical dimensions for this class of models should be $d_c =
\infty$ and that $\tau_{\rm aval}(d = \infty) = 2$, again in good
agreement with (\ref{tau7}).  Those calculations also lead to the
estimates: $\alpha = 0$, $\beta = 0$, $\nu_{\perp} = 0$, and
$\nu_{\parallel} = 1/4$.

  However, from the MP relation (\ref{tau7}), we can derive (see
Appendix A)
\begin{equation}
\gamma_p^{MP} = 1 + \nu_{\perp}
\label{gmp}
\end{equation}
where $\gamma_p$ is the percolation exponent characterizing the
divergence of the average volume of the clusters at the transition.
In Table \ref{expaval}, we show the values obtained for $\gamma_p$ in
our simulations.  We can see that several values are smaller than one,
while (\ref{gmp}) predicts $\gamma_p \ge 1$ for all $d$.

  A limitation of the SOD model is that the velocity of the interface
is constant, thereby making the ``temporal'' definition of avalanches
ambiguous, since our ``physical sense'' of an avalanche is the very
fast invasion of a certain region while here everything happens at
constant velocity \cite{multi}.  To avoid this ``paradox'', a new time
unit can be defined in which the time required to invade a cell would
depend exponentially on the value of $\eta_{\rm min}$.  If we define
this dependence as
\begin{equation}
t \sim \exp \left ( \frac{\eta_{\rm min}}{\eta_o} \right ),
\label{eq111}
\end{equation}
where $\eta_o$ is some constant much smaller than $1$.

\section{The Gradient DPD Model}

  The DPD model can predict the roughness exponent obtained in the
experiments, but several other experimental findings are not explained.
One problem is how to explain that in the experiments the driving force,
at some stage, takes its critical value, allowing the interface to
become pinned.  Also, the effect of evaporation on both the saturation
width $W_{\rm sat}$ and the average height $h_c$ of the pinned interface
cannot be explained by the DPD model.  In order to try to quantify these
ideas, we propose and study in this section an extension to the DPD
model that takes into account these effects.

\subsection{Motivation and Definition}

  We anticipate, on physical grounds, that the smaller is the
evaporation rate \cite{expl1}, the larger will be the critical height.
To test this hypothesis we repeated our imbibition experiments in
different conditions of evaporation.  We find that as we decreased the
evaporation rate, the height reached by the interface indeed increased
(Fig. \ref{exper}).

  The basic idea is that the driving force is no longer a constant, but
rather it depends on the height, so $F = F(h)$. This implies that the
density of blocked cells will also be a function of the height, so $p =
p(h)$.

  In $(1+1)$ dimensions, we model the pinning obstacles by randomly
``blocking'', in a lattice of horizontal size $L$, a fraction $p(h)$ of
the cells in each horizontal row, where $h$ is the height from the
bottom of the lattice, and $p(h)$ is a monotonically increasing function
of $h$.  The original DPD model corresponds to the case $p(h) = const$.

  This extension of the model can be justified on physical grounds.  In
fact, we know that the actual disorder in the paper is {\it not height
dependent}, however, its {\it effect\/} in pinning the propagation of
the fluid is {\it increasing with height}, due to the decrease in the
fluid pressure.  The most physical assumption is an exponential decrease
of the fluid pressure or, equivalently, of the driving force.  This will
lead to an ``effective'' increase in the density of pinning obstacles
\cite{expl2} as we depart from the reservoir.  Hence
\begin{equation}
p(h) - p_o \propto 1 - e^{-h/h_o}.
\label{ppp}
\end{equation}
If $h \ll h_o$ (and $p_c - p_o \ll p_c$), we can write
\begin{equation}
p(h) - p_o \propto h_o^{-1} h \propto gh.
\label{p}
\end{equation}
Hence, in this limit, we find a constant {\it non-zero\/} gradient $g$
in the density of pinning obstacles.

  The gradient in our model is intended to reproduce the combined effect
on the driving force of all the factors referred to in the previous
section: decrease of fluid density because of evaporation, changes in
the density of the suspension, etc.

\subsection{Simulations}

  The presence of the gradient $g$ changes the width of the pinned
interface and its scaling form.  Our simulations show that for
observation scales $\ell$ much smaller than some characteristic
crossover length $\ell_\times'$, the saturated width behaves as $w
\sim \ell^\alpha$, but for $\ell \gg \ell_\times'$, the width
saturates at a value $W_{\rm sat}$ that depends upon the gradient as
\begin{equation}
W_{\rm sat} \sim g^{-\gamma} \qquad [\ell \gg \ell_{\times}'].
\label{w1}
\end{equation}
This behavior can be expressed by a scaling law of the form
\begin{mathletters}
\begin{equation}
w(\ell, g) \sim \ell^{\alpha}~ f_4({\ell}/\ell_\times')
\end{equation}
where
\begin{equation}
\ell_\times' \sim {g^{-{\gamma}/{\alpha}}}.
\end{equation}
\label{w2}
\end{mathletters}
The scaling function $f_4(u)$ satisfies $f_4(u \ll 1) \sim const$ and
$f_4(u\gg 1) \sim u^{- \alpha}$.  Our simulations for a system of size
$L=16~384$ yield the exponents
\begin{equation}
\alpha_{\rm sim} = 0.63 \pm 0.02, \quad \gamma_{\rm sim} = 0.52 \pm 0.02
\qquad [d = 1].
\label{ex1}
\end{equation}
We stress that the validity of the scaling law (\ref{w2}) and the
values of the exponents do not depend on the exact form of $p(h)$ but
only on the value of $\nabla p(h)$ at $h_c$ \cite{Sapoval}.

  The exponent $\gamma$ can be related to $\nu_{\perp}$.  A point of the
interface, at a distance $W_{\rm sat}$ away from the critical height, is
pinned by a DP path if the transverse size of that cluster is of order
$\xi_{\perp}(p)$.  At that point we have $p=p(h_c \pm W_{\rm sat})
\approx p_c \pm g W_{\rm sat}$.  Therefore, using Eq. (\ref{xi}) we
find \cite{Sapoval,wilk,birol,hansen}
\begin{equation}
W_{\rm sat} \sim \xi_{\perp}(p) \sim |p_c - (p_c \pm g W_{\rm
sat})|^{-\nu_{\perp}},
\end{equation}
and
\begin{equation}
W_{\rm sat} \sim |g W_{\rm sat}|^{-\nu_{\perp}}.
\label{gama3}
\end{equation}
  From (\ref{w1}) and (\ref{gama3}) follows
\begin{equation}
\gamma = {\nu_{\perp}}/({1 + \nu_{\perp}}).
\label{gama2}
\end{equation}
Since $\nu_{\perp}$ is known accurately, Eq. (\ref{gama2}) predicts
\begin{equation}
\gamma = 0.523 \pm 0.001 \qquad [d = 1],
\label{eq115}
\end{equation}
in good agreement with our simulation result (\ref{ex1}).

  The generalization of the model, with the gradient, to $(2+1)$
dimensions is straightforward.  We simulated the model for a $512
\times 512$ system; the critical exponents that give the best data
collapse are (see Fig. \ref{gradnum})
\begin{equation}
\alpha_{\rm sim} = 0.43 \pm 0.04 , \quad \gamma_{\rm sim}
= 0.32 \pm 0.02 \qquad [d = 2].
\label{ex2}
\end{equation}
 From these results, we calculate the exponents characterizing the
parallel and perpendicular correlation lengths for the DS problem in
$(2+1)$ dimensions, obtaining
\begin{equation}
\nu_{\parallel} = 1.1 \pm 0.1, \quad  \nu_{\perp} = 0.47 \pm 0.04
\qquad [d = 2].
\label{ex3}
\end{equation}
These results are in reasonable agreement with calculations of the
exponents from the scaling properties of avalanches (see Sec. IV).

\subsection{Experiments}

  Without the gradient, the interface has critical behavior only if we
tune $p_o$ to $p_c$. However, with the gradient the interface always
stops at the critical height $h_c$.  This critical height can be
calculated from the condition $p(h_c)=p_c$.  From (\ref{p}) we obtain
\begin{equation}
h_c \sim g^{-1},
\label{h_c}
\end{equation}
i.e. the height reached by the wetting fluid is inversely proportional
to the gradient in the disorder.

  The experimental data presented in Fig. \ref{gradexp} remarkably
resembles the data obtained from the model.  However, without knowing
the actual value of the gradient in the experiments, it is not
possible to check the validity of the scaling law (\ref{w2})
experimentally.  Nonetheless, measuring the critical height in the
experiments and using Eq. (\ref{h_c}), we are able to estimate $g$,
the gradient in the ``effective disorder'', for the experiments, up to
a multiplicative constant $g_o$.  Using these experimentally
determined values of $g$, we rescale the results obtained for the
width according to the scaling law (\ref{w2}).  In Fig. \ref{gradexp}
we show this rescaling, where we used
\begin{equation}
\alpha_{\rm exp}=0.65 \pm 0.05 , \quad \gamma_{\rm exp}=0.49 \pm 0.05
\qquad [d = 1],
\label{exexp}
\end{equation}
The experimental values of both exponents agree well with the results
obtained from the simulations (Table \ref{expgrad}) and with the
theoretical predictions based in the known results from DP.

  In summary, in this section we discussed a variant of the DPD model
that incorporates evaporation by introducing a gradient in the density
of pinning cells \cite{expl2}. The model provides insight into three
previously-unexplained aspects of imbibition experiments:

  {\it (i)\/} The interface always stops growing, after some finite
time.  Due to the gradient, the wetting interface only moves until it
reaches a critical density of pinning cells. This gradient in pinning
cells arises from the balance between the evaporation of the fluid and
the surface tension in one hand, and the capillary forces tending to
move it along the paper in the other.

  {\it (ii)\/} The final height of the interface, $h_c$, increases when
the evaporation is reduced, due to the smaller {\it effective\/}
gradient in the pinning disorder.

  {\it (iii)\/} A exponent $\gamma$ was found characterizing the
dependence on the gradient of the saturation width and the
characteristic length $\ell_\times$.  Good agreement was found between
experimental, theoretical and numerical calculations of the exponents.

\section{Summary}

  In this work, we introduced and discussed a set of imbibition
experiments that probe the behavior of rough interfaces in disordered
media.  We developed a discrete model, the DPD model, and showed that
it correctly predicts the experimental results.

  We presented a discussion of the properties of the DPD model and
argued that it can be mapped to a continuum differential equation of the
KPZ-type with quenched disorder.  We showed that the model has only
three independent exponents --- $\nu_{\parallel}$, $\alpha$, and $z$ ---
from which all others can be derived through scaling laws.  The mapping
to DP enables us to obtain $\nu_{\parallel}$ and $\alpha$ from the
exponents of DP ($d = 1$) or DS ($d > 1$).  On the other hand, a mapping
of the dynamics of the DPD model to {\it isotropic\/} percolation allow
us to determine $z$.

  We introduced avalanches in the DPD model, and studied their scaling
properties.  We then related the avalanches in the DPD model to the
avalanches in the SOD variant of the DPD model.  We derive scaling laws
relating the critical exponents for the avalanches, $\tau_{\rm aval}$
and $\tau_{\rm surv}$, to the other critical exponents of the DPD model
confirming that they are not independent.

  Finally, we performed a new set of imbibition experiments to study
the effect of evaporation on interfacial phenomena.  We modify the DPD
model to take these effects in consideration and use the new model to
predict the experimental results.  Again the mapping to DP enables
us to estimate the values of the exponents.  Good agreement was
obtained between experimental, theoretical and numerical calculations.

\acknowledgments

  We thank I.~S. Buldyrev, R. Cuerno, J. Kert\'esz, K.~B.  Lauritsen, H.
Makse, S. Schwarzer, M. Ukleja, T. Vicsek, and P.-z. Wong for valuable
contributions and discussions.  LANA acknowledges a fellowship from
Junta Nacional de Investiga\c c\~ao Cient\'{\i}fica e Tecnol\'ogica.  SH
acknowledges partial support from the Bi--National US--Israel Foundation
and the Minerva Center for Mesoscopic Physics, Fractals and Neural
Networks. The Center for Polymer Studies is supported by the National
Science Foundation.

\appendix
\section{Calculating the Correlation Length Exponents}

  In this appendix we discuss the relation between percolation
and the growth of clusters by avalanches, and use the insight gained
to determine the exponents characterizing the divergence of the
parallel and perpendicular correlation lengths.

  As discussed above, the characteristic volume of the avalanches
diverges when $p \rightarrow p_c$,
\begin{equation}
s_o \sim (p_c - p)^{-1/ \sigma},
\label{eq121}
\end{equation}
where here \cite{perc1,perc2}
\begin{equation}
1/ \sigma = d \nu_{\parallel} + \nu_{\perp}.
\label{ap1}
\end{equation}
The average cluster size also diverges for the critical
probability
\begin{equation}
\langle s \rangle \equiv \int P_{\rm aval}(s) ds \sim (p_c -
p)^{- \gamma_p},
\label{eq122}
\end{equation}
where
\begin{equation}
\gamma_p = \frac{2 - \tau_{\rm aval}}{\sigma}.
\label{ap2}
\end{equation}
In percolation theory \cite{perc1,perc2} this exponent is called
$\gamma$.  To avoid confusion with the growth exponent $\beta$, we
denote the exponent characterizing the mass of the infinite cluster as
$\beta_p$, where
\begin{equation}
\beta_p = 1 / \sigma - \gamma_p.
\label{ap3}
\end{equation}

  We calculated both $\tau_{\rm aval}$ and $\sigma$ from the
scaling of $P_{\rm aval}(s)$ for $d = 1, 2$.  We obtained
\cite{Buldyrev1,Buldyrev2}
\begin{mathletters}
\begin{equation}
\tau_{\rm aval} = 1.26\pm 0.02, \quad 1/\sigma = 2.8\pm 0.1
\qquad [d = 1],
\end{equation}
and
\begin{equation}
\tau_{\rm aval} = 1.51\pm 0.07, \quad 1/\sigma = 2.9\pm 0.3
\qquad [d = 2].
\end{equation}
\label{tau1}
\end{mathletters}
In $(1+1)$ dimensions the value of $\sigma$ is in good agreement with
the predictions of DP, $1/\sigma \simeq 2.83$.  However, the values of
$\gamma_p$ and $\beta_p$ differ from the predictions of DP.  This is
not surprising since the definitions of clusters in both models are
quite different.  For the DPD model the clusters are the avalanches
(compact regions confined between two DP paths), while the DP cluster
are branchy trees comprised of directed paths.

  In percolation theory, the correlation lengths are defined through
the formulas
\begin{mathletters}
\begin{equation}
\xi_{\perp}^2 \equiv \frac{\langle s R_{\perp}^2 \rangle}
{\langle s \rangle} \sim (p_c - p)^{-2 \nu_{\perp}},
\end{equation}
\begin{equation}
\xi_{\parallel}^2 \equiv \frac{\langle s R_{\parallel}^2 \rangle}
{\langle s \rangle} \sim (p_c - p)^{-2 \nu_{\parallel}},
\end{equation}
\label{xidef}
\end{mathletters}
\noindent
where the brackets denote an average of realizations of the disorder,
and $R_{\perp}^2$ and $R_{\parallel}^2$ are the perpendicular and
parallel components of the square of the radius of gyration of the
avalanches,
\begin{mathletters}
\begin{equation}
R_{\perp}^2 \equiv \overline{h^2} - {\overline{h}}^2,
\end{equation}
\begin{equation}
R_{\parallel}^2 \equiv \overline{x^2} - {\overline{x}}^2,
\end{equation}
\label{r2def}
\end{mathletters}
\noindent
where the bar denotes a spatial average for all cells of the avalanche
in a given realization of the disorder.  The relations (\ref{xidef})
and (\ref{r2def}) were applied to the calculation of the correlation
length exponents presented in Sec. IV.  The results obtained are in
good agreement with theoretical calculations for DP in $(1+1)$
dimensions.

\section{Calculating the Velocity Exponent}

  The purpose of this appendix is to propose an alternative method for
the calculation of the velocity exponent $\theta$.  The traditional
way to calculate $\theta$ is to monitor the average height of the
interface after the system becomes saturated.  This is a very time
consuming procedure for two reasons.  First, the dependence of the
saturation time $t_{\times}$ on the system size, $t_{\times} \sim
L^z$, leads to an great increase in computation time when we use large
system sizes.  Second, the large fluctuations in the velocity of about
its average value leads to an increase in the number of runs required
for achieving good statistics.  The reason for the large fluctuations
of the velocity are due to the fact that, near the transition, the
motion of the interface is not smooth but rather is very ``jerky'',
i.e. short periods of jumpy motion are followed by long periods of
near immobility.

  We assume that, before saturation, the velocity scales with time
as a power law with an exponent $\psi$
\begin{equation}
v(t,f) \sim t^{-\psi} f_5 (t / t_{\times}),
\label{vvv}
\end{equation}
with $t_{\times} \sim \xi_{\parallel}^z$.  Comparison of (\ref{vvv})
with (\ref{vel}) implies that $f_5(u)$ is a scaling function that
satisfies $f_5(u \ll 1) = const$, and $f_5(u \gg 1) \sim u^{\psi}$,
and that
\begin{equation}
\theta = \nu_{\parallel}~z~\psi.
\label{psi}
\end{equation}
The calculation of the exponents $\psi$, $z$ and $\nu_{\parallel}$
then allows us to estimate $\theta$.  Numerical tests reveal that
(\ref{psi}) is well obeyed.

  Relation (\ref{psi}) can be combined with (\ref{thet}) leading to
the scaling law
\begin{equation}
\psi = 1 - \beta.
\end{equation}
This new relation provides an alternate method to check the
consistency of the scaling laws derived in the text.

\begin{figure}
\narrowtext
\caption{  The depinning transition.  In the ``pinned phase'' (pp), $F <
F_c$, the velocity of the interface is zero.  In the ``moving phase''
(mp), $F > F_c$, the interface moves with a constant average velocity
$v$.  The velocity plays the role of the order parameter of the
transition.  }
\label{phases}
\end{figure}

\begin{figure}
\narrowtext
\caption{  (a) Schematic illustration of the experimental setup.  In the
$(1+1)$ dimensional experiments we use paper towels as the disordered
media and coffee as the invading fluid.  The edge of the paper towels
is 20 cm.  (b) Photographs of pinned interfaces in imbibition
experiments with coffee and paper towels for (i) high evaporation
rate: $g_{exp} = 0.94 g_o$, and (ii) low evaporation rate: $g_{exp} =
0.25 g_o$. Here $g_o$ is the undetermined multiplicative constant
discussed in Sec. V.}
\label{exper}
\end{figure}

\begin{figure}
\narrowtext
\caption{  Experimental local width $w$ for the pinned and moving
interfaces in $(1+1)$ dimensions.  It is visually apparent that the
moving interface has a larger width and a different roughness exponent
than the pinned interface.}
\label{wexp2d}
\end{figure}

\begin{figure}
\narrowtext
\caption{  Digitized ink interface in the (a) ``Oasis'' brick, and (b)
paper roll using a Apple Scanner with a resolution of 300 pixels per
inch.  In (c) we show the full image from which (b) was magnified.
The brick has a section of $7 \times 7$ square centimeters, and the
paper roll an exterior radius of $7.5$ centimeters and an interior
radius of $1.75$ centimeters.}
\label{exp2pic}
\end{figure}

\begin{figure}
\narrowtext
\caption{  Scaling of the local width $w$ with $\ell$ for the
experiments in $(2+1)$ dimensions .  The curve for the paper roll
results from averaging over 13 different paper sheets, and the one
for the ``Oasis'' brick results from averaging over 9 different
sections.  }
\label{exp2wid}
\end{figure}

\begin{figure}
\narrowtext
\caption{Example of the application of the growth rule to a particular
configuration of the interface.  In the figures, white squares refer
to dry unblocked cells, the darker squares refer to dry blocked cells,
and the gray squares to wet cells.  The tick line shows the position of
the interface.  Let us suppose that the column indicated by the arrow
in (a) was chosen for growth.  According to our model, the cells
marked '1' will become wet because they are nearest-neighbors to the
wet region.  After the wetting, we can see in (b) that the cell marked
'2' is below a wet cell, applying the rule to erode any overhang we
wet that cell.  In (c) we show the configuration of the interface
after the growth (thick line).}
\label{model}
\end{figure}

\begin{figure}
\narrowtext
\caption{General conditions for the pinning of the interface.  We
see in the figure that the path of blocked cells pinning the interface
is connected as a DP path, with 5 possible directions (North, Northeast,
East, Southeast, and South).  The shadowing convention for the cells is
the same as in Fig. 6.}
\label{pinmodel}
\end{figure}

\begin{figure}
\narrowtext
\caption{Values of the exponents $\alpha$ and $\beta$, for the DPD
model for dimensions up to $(6+1)$.  The results plotted suggest that
this class of models has no critical dimension.}
\label{dpalp}
\end{figure}

\begin{figure}
\narrowtext
\caption{(a) Initial conditions in the DPD model for the growth of
single avalanches: all sites in the bottom edge except one are
blocked.  The color convention for the cells is the same as in Fig. 1.
(b) The wet region after some time, the quantity $n(t)$ defined in the
text counts the number of unblocked cells in the interface, in this
case 8.  (c) Horizontal projection of an avalanche in $(2+1)$
dimensions for $p \simeq p_c$.  The avalanche was started at the
center of the figure and is shown at time $2^{10}$. The current
diameter of the cluster is approximately $2^{10}$.  The uniform gray
area shows the region left dry since the beginning of the process. The
darkest shades of gray corresponds to the largest heights of the
interface. The black dots, forming a ``fractal dust'', indicate the
unblocked cells at the interface.}
\label{avalproj}
\end{figure}

\begin{figure}
\narrowtext
\caption{  Illustration of the dynamics of the DPD model for $(1+1)$
dimensions. (a) {\it Schematic representation \/} of a region defined
by two pinning paths.  The heavy circle indicates the origin for the
invasion, the thin arcs represent the positions of the invading front
at successive times, and the dashed line represents schematically the
path for the invasion.  (b) {\it Simulation \/} results for invasion
after $2^{10}$ time steps starting from a single cell near the center.
We show the invaded region at a sequence of times which are multiples
of $128$.  Regions invaded at later times are displayed in darker
shades of gray.  The path from the origin to the latest invaded point
is shown in black.  Although this path displays some fluctuations in
the vertical direction, they can be disregarded since $\nu_{\parallel}
> \nu_{\perp}$, so as $p \rightarrow p_c$, $\xi_{\perp} /
\xi_{\parallel} \rightarrow 0$.  Thus the distance propagated by the
invading front is proportional to time.  Since $t_{\times} \sim \ell$,
we can conclude that $z = d_{\rm min} = 1$.  }
\label{tx1}
\end{figure}

\begin{figure}
\narrowtext
\caption{  Illustration of the dynamics of the DPD model for $(2+1)$
dimensions. (a) {\it Schematic representation \/} of the $xy$
projection of the region defined by two pinning self-affine DS.  The
heavy circle indicates the origin for the invasion, the thin arcs
represent the $xy$ projections of the invading front at successive
times, and the dashed line represents schematically the path for the
invasion.  (b) {\it Simulation \/} results for invasion after $2^{10}$
time steps starting from a single cell located to the left of the
center.  We show the $xy$ projection of the invaded region at a
sequence of times which are multiples of $128$.  Regions invaded at
later times are displayed in darker shades of gray.  It is visually
apparent that it takes a long time to invade some regions close to the
origin because the path to that position (shown in black) appears to
be a fractal curve of dimension greater than one.  The fluctuations in
the vertical direction can be disregarded since we know that
$\xi_{\perp} / \xi_{\parallel} \rightarrow 0$.  We find that the path
can be identified with the shortest path (the ``chemical distance'')
of isotropic percolation, and that its length scales with the linear
distance $r$ to the point as $r^{d_{\rm min}}$.  }
\label{tx2}
\end{figure}

\begin{figure}
\narrowtext
\caption{ (a) Scaling with time of the horizontal length of a DPD
cluster in $(d+1)$ dimensions grown from a single cell.  Shown is a
plot of time $t$ as a function of $r_{\parallel}$, which is the
average of the parallel components of the radius of gyration of the
shell.  The asymptotic slope is $z$.  (b) Consecutive slopes analysis
of the data displayed in (a).  Note that, after some transient
behavior, a transition to a power law scaling occurs.  For higher
dimensions, the power law scaling is affected by finite-size effects
for larger times.  }
\label{zscal}
\end{figure}

\begin{figure}
\narrowtext
\caption{Successive series of pinned interfaces for the DP model with
$L=400$, $p = 0.5 > p_c$.   We show the boundaries of avalanches,
produced by removing a randomly-chosen blocked cell from the
previously-pinned interface.  The correlation lengths $\xi_{\parallel}$
and $\xi_{\perp}$ displayed describe the typical size of the
avalanches.}
\label{aval}
\end{figure}

\begin{figure}
\narrowtext
\caption{Distribution of survival times for avalanches $P_{\rm surv}(t)$
for $(2+1)$ dimensions.  The system size is $2~048$ and $10^6$ avalanches
were produced for each probability.  In (a) we show data for several
probabilities close to the critical value.  The straight part of the
distribution can be well fitted to a power law with exponent $2.18$.  In
(b) we show the consecutive slopes of the data in (a).  The figure makes
clear that we can predict with great precision the value of $p_c$ and
the error of the estimate.}
\label{avaltau}
\end{figure}

\begin{figure}
\narrowtext
\caption{Distribution of avalanches size $P_{\rm aval}$ for $(2+1)$
dimensions.  The system size is $2~048$ and $10^6$ avalanches were
produced for each probability.  In (a) we show the data for the critical
probability, a good power law fit can be done with an exponent $1.51$.
(b) shows a data collapse, according to (4.2), for several distinct
probabilities close to $p_c$.}
\label{avalt1}
\end{figure}

\begin{figure}
\narrowtext
\caption{Scaling of the average size $s$, and parallel and perpendicular
correlation lengths of the avalanches for $(2+1)$ dimensions.  These
quantities were calculated according to the method described in
Appendix A.  Good power law fitting was found with the exponents given
in (4.16).}
\label{avalnu}
\end{figure}

\begin{figure}
\narrowtext
\caption{  Evolution with time of $\eta_{\rm min}$.  Every local
maxima defines the end of an avalanche and the beginning of another.
Regions with monotonic growth indicate that the avalanches have size
one.  After the transient period, the critical state is reached and
the maxima become equal to $q_c = 1 - p_c$.}
\label{min}
\end{figure}

\begin{figure}
\narrowtext
\caption{The simulation results for the width $w(\ell,g)$ of the
pinned interface in $(2+1)$ dimensions, where $g$ is the gradient in the
density of blocked cells. (a) The widths for several values of the
gradient (averaged over 128 runs for each value of the gradient).  (b)
The same simulation results, plotted in the scaling form of Eq. (5.4),
using the values of the exponents from (5.10).  }
\label{gradnum}
\end{figure}

\begin{figure}
\narrowtext
\caption{The experimental results for the width  $w(\ell,g)$ of
the pinned interface.  (a) The widths for several values of the gradient
$g$ (in units of $g_o$). The values of $g$ were calculated as described
in the text; the error in these values is smaller than $10\%$.  The
widths were corrected by a multiplicative factor to make them coincide
for the smallest $\ell$. (b) The same experimental results, plotted in
the scaling form of Eq. (5.4), using the values of the exponents from
(5.13). }
\label{gradexp}
\end{figure}

\end{multicols}

\begin{table}
\caption{Critical exponents of the DPD model for dimensions up to
$(6+1)$ as measured directly from the simulations.  }
\begin{tabular}{c|cccccc}
Exponents     & $(1+1)$ & $(2+1)$       & $(3+1)$       & $(4+1)$
        & $(5+1)$       & $(6+1)$       \\
\tableline
$\nu_{\parallel}$ & $1.73\pm 0.02$   & $1.16\pm 0.05$    & $0.95\pm 0.1$
        & $0.66\pm 0.10$        & $0.6\pm 0.1$        & $0.5\pm 0.1$ \\
$\alpha$          & $0.63\pm 0.01$   & $0.48\pm 0.03$    & $0.38\pm 0.04$
        & $0.27\pm 0.05$        & $0.25\pm 0.05$      & $0.2\pm 0.2$ \\
$z$               & $1.01\pm 0.02$   & $1.15\pm 0.05$    & $1.36\pm 0.05$
        & $1.58\pm 0.05$        & $1.7\pm 0.1$        & $1.8\pm0.2$  \\
\tableline
$\nu_{\perp}$     & $1.10\pm 0.02$   & $0.57\pm 0.05$    & $0.34\pm 0.05$
        & $0.2\pm 0.1$         & $0.15\pm 0.05$       & $0.1\pm 0.1$ \\
$\beta$           & $0.63\pm 0.01$   & $0.41\pm 0.05$    & $0.29\pm 0.07$
        & $0.2\pm 0.1$         & $0.1\pm 0.1$         & $0.1\pm 0.1$ \\
$\psi$            & $0.33\pm 0.04$   & $0.59\pm 0.07$    & $0.74\pm 0.09$
        & $1.00\pm 0.09$        & $$                  & $$ \\
$\theta$          & $0.58\pm 0.07$   & $0.8\pm 0.2$      & $1.0\pm 0.2$
        & $1.0\pm 0.2$          & $$                  & $$ \\
\end{tabular}
\label{expdim}
\end{table}

\begin{table}
\narrowtext
\caption{ Dynamical exponent $z$ for the DPD model in $(d+1)$
dimensions and the shortest path exponent $d_{\rm min}$ for isotropic
percolation for a $d$-dimensional cubic lattice of $L^d$ sites.  The
values indicated by an asterisk are exact, while the remaining values
were calculated in our simulations by studying the consecutive slopes of
the linear regime in Fig. 12.  At the critical dimension $d_c = 6$, one
should not expect to find the exact result $d_{\rm min} = 2$ because
logarithmic corrections are generally present.  The system sizes used in
the simulations range from $L = 4~096$ for $d = 2$, to $L = 16$ for $d
= 6$.  }
\begin{tabular}{c|cc|cc}
  & \multicolumn{2}{c}{DPD} &  \multicolumn{2}{c}{Percolation} \\
$~~d~~$    & $p_c$              & $z$            & $p_c$
        & $d_{\rm min}$ \\
\tableline
$~1~$      & $0.4698\pm 0.0005$ & $1.01\pm 0.02$ & $1^{ *}$
        & $1^{ *}$        \\
$~2~$      & $0.7425\pm 0.0005$ & $1.15\pm 0.05$ & $0.5927\pm 0.0005$
        & $1.13\pm 0.03$  \\
$~3~$      & $0.8425\pm 0.0005$ & $1.36\pm 0.05$ & $0.3116\pm 0.0005$
        & $1.38\pm 0.02$  \\
$~4~$      & $0.8895\pm 0.0005$ & $1.58\pm 0.05$ & $0.197\pm 0.005$
        & $1.53\pm 0.05$  \\
$~5~$      & $0.9175\pm 0.0005$ & $1.7\pm 0.1$   & $0.141\pm 0.005$
        & $1.7\pm 0.1$   \\
$~6~$      & $0.931\pm 0.005$   & $1.8\pm0.2$    & $0.107\pm 0.005$
        & $1.8\pm 0.2$   \\
\end{tabular}
\label{tab1}
\end{table}

\begin{table}
\narrowtext
\caption{  Critical exponents for the avalanches in the DPD model for
dimensions up to $(6+1)$ as measured directly from the simulations.
For comparison we also show the estimates obtained with (4.17) and
(4.18).  }
\begin{tabular}{c|cccccc}
Exponents     & $(1+1)$ & $(2+1)$       & $(3+1)$       & $(4+1)$
        & $(5+1)$       & $(6+1)$       \\
\tableline
$\gamma_p$                      & $1.98\pm 0.03$   & $1.41\pm 0.05$
        & $0.95\pm 0.05$        & $0.8\pm 0.2$     & $0.5\pm 0.2$
        & $0.3\pm 0.2$\\
$\delta$                        & $0.60\pm 0.03$   & $1.14\pm 0.06$
        & $1.6\pm 0.1$          & $1.9\pm 0.2$     & $2.1\pm 0.2$
        & $2.5\pm 0.3$ \\
$\tau_{\rm surv}$               & $1.46\pm 0.02$   & $2.18\pm 0.03$
        & $2.54\pm 0.05$        & $3.0\pm 0.2$     & $ $
        & $ $ \\
$\tau_{\rm aval}$               & $1.26\pm 0.02$   & $1.51\pm 0.07$
        & $1.70\pm 0.05$        & $1.7\pm 0.1$     & $1.8\pm 0.1$
        & $1.9\pm 0.1$ \\
$\tau_{\rm aval}^{\rm MP}$      & $1.3\pm 0.1$     & $1.5\pm 0.1$
        & $1.6\pm 0.2$          & $1.6\pm 0.2$     & $1.6\pm 0.1$
        & $1.6\pm 0.1$ \\
$\tau_{\rm aval}^{\rm OPZ}$     & $1.2\pm 0.1$     & $1.2\pm 0.1$
        & $1.2\pm 0.1$          & $1.2\pm 0.1$     & $1.1\pm 0.1$
        & $1.1\pm 0.1$ \\
\end{tabular}
\label{expaval}
\end{table}

\begin{table}
\caption{Critical exponents of the gradient DPD model.  }
\begin{tabular}{c|cc}
Exponents               & $(1+1)$         & $(2+1)$ \\
\tableline
$\alpha$                & $0.63 \pm 0.02$ & $0.43 \pm 0.04$ \\
$\gamma$                & $0.52 \pm 0.02$ & $0.32 \pm 0.02$ \\
$\nu_{\perp}$           & $1.09 \pm 0.08$ & $0.47 \pm 0.04$ \\
$\nu_{\parallel}$       & $1.7 \pm 0.1$   & $1.1 \pm 0.1$ \\
\end{tabular}
\label{expgrad}
\end{table}


\begin{references}
\bibitem{surf}
Smaller pinning regions might become totally surrounded by the wet
region.  However, because of the surface tension that acts to decrease
the size of the interface between the wet and dry regions, these dry
``islands'' inside the wet region eventually become wet.
\bibitem{expl0}
We introduce this time-dependent randomness in the model to mimic the
experimental fact that for $F > F_c$ a transition to the KPZ behavior
with annealed disorder is observed for $\ell \gg \xi$.  For smaller
length scales the effect of this annealed randomness is irrelevant.
\bibitem{explain0}
The justification of this particular rule lies not only on the
computational simplifications that it affords but also on the
experimental finding that overhangs do not play an important role in the
scaling properties of the interface for imbibition.
\bibitem{fitz}
We found that the values of $z$ could be well approximated by the
mnemonic formula: $z = 2 - [(6-d)/5]^{3/4}$.  Although this expression
is uncompatible with an $\epsilon$ expansion, and so cannot be correct,
it can be useful in the estimation of the exponents.
\bibitem{rule}
For the study of the avalanches it is more convenient to
consider a new definition of the unit time in which, for each time
step all unblocked nearest neighbors to the invading region also
become invaded.  This means that we make a parallel updating of the
system.  For $p$ close to $p_c$, this produces no changes in the
results, however for $p \ll p_c$ this parallel updating leads to an
essentially flat interface.
\bibitem{expl1}
In this work we refer to evaporation as the origin of the gradient.
However, other factors also affect the value of the experimental
gradient, among them the density of the suspension. Indeed, our
experiments confirm this idea and suggest that the model accounts
for all these effects.
\end{references}
\end{document}